\documentclass{elsart5p}
\usepackage{natbib}
\usepackage{graphicx}
\usepackage{amssymb}
\usepackage{mathrsfs} 

\newtheorem{theorem}{Theorem}%[section]
\newtheorem{corollary}[theorem]{Corollary} 

\newtheorem{proposition}[theorem]{Proposition}
\theoremstyle{definition}

\theoremstyle{remark}
\newtheorem{remark}[theorem]{Remark}

\newtheorem{example}[theorem]{Example}

\def\norm#1{\left\Vert#1\right\Vert}

\def\R{{\mathbf R}}

\def\e{\varepsilon}
\newcommand{\s}{\mathbb{S}}
\newcommand{\abs}[1]{\vert#1\vert}
\newcommand{\sep}{\mathrm{sep}}

\begin{document}

\begin{frontmatter}

\title{An axiomatic approach to intrinsic dimension of a dataset}

\author{Vladimir Pestov}

\address{Department of Mathematics and Statistics, University of Ottawa,
585 King Edward Avenue, Ottawa, ON, Canada K1N 6N5}

\begin{abstract}
We perform a deeper analysis of an axiomatic approach to the concept of intrinsic dimension of a dataset proposed by us in the IJCNN'07 paper.
The main features of our approach are that a high intrinsic dimension of a dataset reflects the presence of the curse of dimensionality (in a certain mathematically precise sense), and that dimension of a discrete i.i.d. sample of a low-dimensional manifold is, with high probability, close to that of the manifold. At the same time,
the intrinsic dimension of a sample is easily corrupted by moderate high-dimensional noise (of the same amplitude as the size of the manifold) and suffers from prohibitevely high computational complexity (computing it is an $NP$-complete problem). We outline a possible way to overcome these difficulties. 
\end{abstract}


\end{frontmatter}

\section{Introduction}
\label{s:introduction}

An often-held opinion on intrinsic dimensionality of data sampled from submanifolds of the Euclidean space 
is expressed in \cite{hein:05} thus:
``...the goal of estimating the dimension of a submanifold is a well-defined mathematical problem. Indeed all the notions of dimensionality like e.g. topological, Hausdorff, or correlation dimension agree for submanifolds in $\R^d$.'' 

We will argue that it may be useful to have at one's disposal a concept of intrinsic dimension of data which behaves in a different fashion from the more traditional concepts. 

Our approach is shaped up by the following five goals.

1. We want a high value of intrinsic dimension to be indicative of the presence of the curse of dimensionality. 

2. The concept should make no distinction between continuous and discrete objects, and
the intrinsic dimension of a discrete sample should be close to that of the underlying manifold. 

3.  The intrinsic dimension should agree with our geometric intuition and return standard values for familiar objects such as Euclidean spheres or Hamming cubes.

4. We want the concept to be insensitive to high-dimensional random noise of moderate amplitude
(on the same order of magnitude as the size of the manifold). 

5. Finally, in order to be useful, the intrinsic dimension should be computationally feasible. 

For the moment, we have managed to attain the goals (1),(2),(3), while (4) and (5) are not met. However, it appears that in both cases the problem is the same, and we outline a promising way to address it. 

Among the existing approaches to intrinsic dimension, that of \cite{chavez:01} comes closest to meeting the goals (2),(3),(5) and to some extent (1), cf. a discussion in \cite{pestov:07}. (Lemma 1 in \cite{hein:07} seems to imply that (4) does not hold for moderate noise with ${\mathbb E}\norm x = O(1)$, i.e., $\sigma^2\sim 1/d$.) 

We work in a setting of metric spaces with measure ($mm$-spaces), i.e., triples $(X,d,\mu)$ consisting of a set, $X$, furnished with a distance, $d$, satisfying axioms of a metric, and a probability measure $\mu$.
This concept is broad enough so as to include submanifolds of $\R^n$ (equipped with the induced, or Minkowski, measure, or with some other probability distribution), as well as data samples themselves (with their empirical, that is normalized counting, measure). In Section \ref{s:conc}, we describe this setting and discuss in some detail the phenomenon of concentration of measure on high dimensional structures, presenting it from a number of different viewpoints, including an approach of soft margin classification.

The curse of dimensionality is understood as a geometric property of $mm$-spaces whereby features ($1$-Lipschitz, or non-expanding, functions) sharply concentrate near their means and become non-discriminating. This way, the curse of dimensionality is equated with the phenomenon of concentration of measure on high-dimensional structures \cite{milman:00,gromov:99}, and can be dealt with an a precise mathematical fashion, adopting (1) as an axiom.

The intrinsic dimension, $\partial$, is defined for $mm$-spaces in an axiomatic way in Section \ref{s:axiom}, following \cite{pestov:07}.

To deal with goal (2), we resort to the notion of a distance, $d_{conc}(X,Y)$, between two $mm$-spaces, $X$ and $Y$, measuring their similarity  \cite{gromov:99}. This forms the subject of Section \ref{s:gromov}.
Our second axiom says that if two $mm$-spaces are close to each other in the above distance, then their intrinsic dimension values are also close. In this article, we show that if a dataset $X$ is sampled with regard to a probability measure $\mu$ on a manifold $M$, then, with high confidence, the distance between $X$ and $M$ is small, and so $\partial(M)$ and $\partial(X)$ are close to each other.

The goal (3) can be made into an axiom in a more or less straightforward way. 
We give a new example of a dimension function $\partial$ satisfying our axioms. 

We show that the Gromov distance between a low-dimensional manifold $M$ and its
corruption by high-dimensional gaussian noise of moderate amplitude is close to $M$ in the Gromov distance. However, this property does not carry over to the samples unless their size is exponential in the dimension of $\R^d$ (unrealistic assumption), and thus our approach suffers from high sensitivity to noise (Section \ref{s:noise}.) Another drawback is computational complexity: we show that computing the intrinsic dimension of a finite sample is an $NP$-complete problem (Sect. \ref{s:complexity}.)

However, we believe that the underlying cause of both problems is the same: allowing {\em arbitrary} non-expanding functions as features is clearly too generous. Restricting the class of features to that of low-complexity functions whose capacity is manageable and rewriting the entire theory in this setting opens up a possibility to use statistical learning theory and offers a promising way to solve both problems, which we discuss in Conclusion. 

\section{\label{s:conc}The phenomenon of concentration of measure on high-dimensional structures}

\subsection{Spaces with metric and measure}

As in \cite{pestov:07}, we model datasets within the framework of spaces with metric and measure ($mm$-spaces). So is called a triple $(X,d,\mu)$, consisting of a (finite or infinite) set $X$, a metric $d$ on $X$, and a  probability measure\footnote{That is, a sigma-additive measure of total mass one.} $\mu$ defined on the family $\mathscr B$ of all Borel subsets\footnote{Recall that $\mathscr B$ is the smallest family of subsets of $X$ closed under countable unions and complements and containing every open ball $B_\e(x)$, $\e>0$, $x\in X$.} of the metric space $(X,d)$.

The setting of $mm$-spaces is natural for at least three reasons. First, a finite dataset $X$ sitting in a Euclidean space $\R^d$ forms an $mm$-space in a natural way, as it
comes equipped with a distance and a probability measure (the empirical measure $\mu_{\sharp}(A)=\sharp(A)/\sharp(X)$, where $\sharp(A)$ denotes the number of elements in $A$). Second, 
if one wants to view datasets as random samples, then 
the domain $\Omega$, equipped with the sampling measure $\mu$ and a distance, also forms an $mm$-space. And finally, theory of $mm$-spaces is an important and fast developing part of mathematics, the object of study of asymptotic geometric analysis, see \cite{milman:86,milman:00,gromov:99} and references therein.

{\em Features} of a dataset $X$ are functions on $X$ that in some sense respect the intrinsic structure of $X$. In the presence of a metric, they are usually understood to be {\em 1-Lipschitz,} or {\em non-expanding,} functions $f$, that is, having the property  
\[\abs{f(x)-f(y)}\leq d(x,y)\mbox{ for all }x,y\in X.\]
We will denote the collection of all real-valued 1-Lipschitz functions on $X$ by ${\mathrm{Lip}}_1(X)$. 

\subsection{Curse of dimensionality and observable diameter}

The curse of dimensionality is a name given to the situation where all or some of the important features of a dataset sharply concentrate near their median (or mean) values and thus become non-discriminating. In such cases, $X$ is perceived as intrinsically high-dimensional. This set of circumstances covers a whole range of well-known high-dimensional phenomena such as for instance sparseness of points (the distance to the nearest neighbour is comparable to the average distance between two points \cite{BGRS}), etc. It has been argued in \cite{pestov:00} that a mathematical counterpart of the curse of dimensionality is the well-known {\em concentration phenomenon} \cite{milman:00,L}, which can be expressed, for instance, using Gromov's concept of the {\em observable diameter} \cite{gromov:99}.

Let $(X,d,\mu)$ be a metric space with measure, and let $\kappa>0$ be a small fixed threshold value. 
The {\em observable diameter} of $X$ is the smallest real number, $D={\mathrm{ObsDiam}}_\kappa(X)$, with the following property: for every two points $x,y$, randomly drawn from $X$ with regard to the measure $\mu$, and for any given $1$-Lipschitz function $f\colon X\to\R$ (a feature), the probability of the event that values of $f$ at $x$ and $y$ differ by more than $D$ is below the threshold:
\[P[\abs{f(x)-f(y)}\geq D]<\kappa.\]
Informally, the observable diameter ${\mathrm{ObsDiam}}_\kappa(X)$ is the size of a dataset $X$ as perceived by us through a series of randomized measurements using arbitrary features and continuing until the probability to improve on the previous observation gets too small. The observable diameter has little (logarithmic) sensitivity to $\kappa$.

The {\em characteristic size} ${\mathrm{CharSize}}\,(X)$ of $X$ as the median value of distances between two elements of $X$. 
The concentration of measure phenomenon refers to the observation that ``natural'' families of geometric objects $(X_n)$ often satisfy
\[{\mathrm{ObsDiam}}_\kappa (X_n)\ll {\mathrm{CharSize}}\,(X_n)\mbox{ as }n\to\infty.\]
A family of spaces with metric and measure having the above property is called a {\em L\'evy family}. Here the parameter $n$ usually corresponds to dimension of an object defined in one or another sense.

\begin{figure}[htp]
\centerline{\scalebox{0.22}[0.271]{\includegraphics{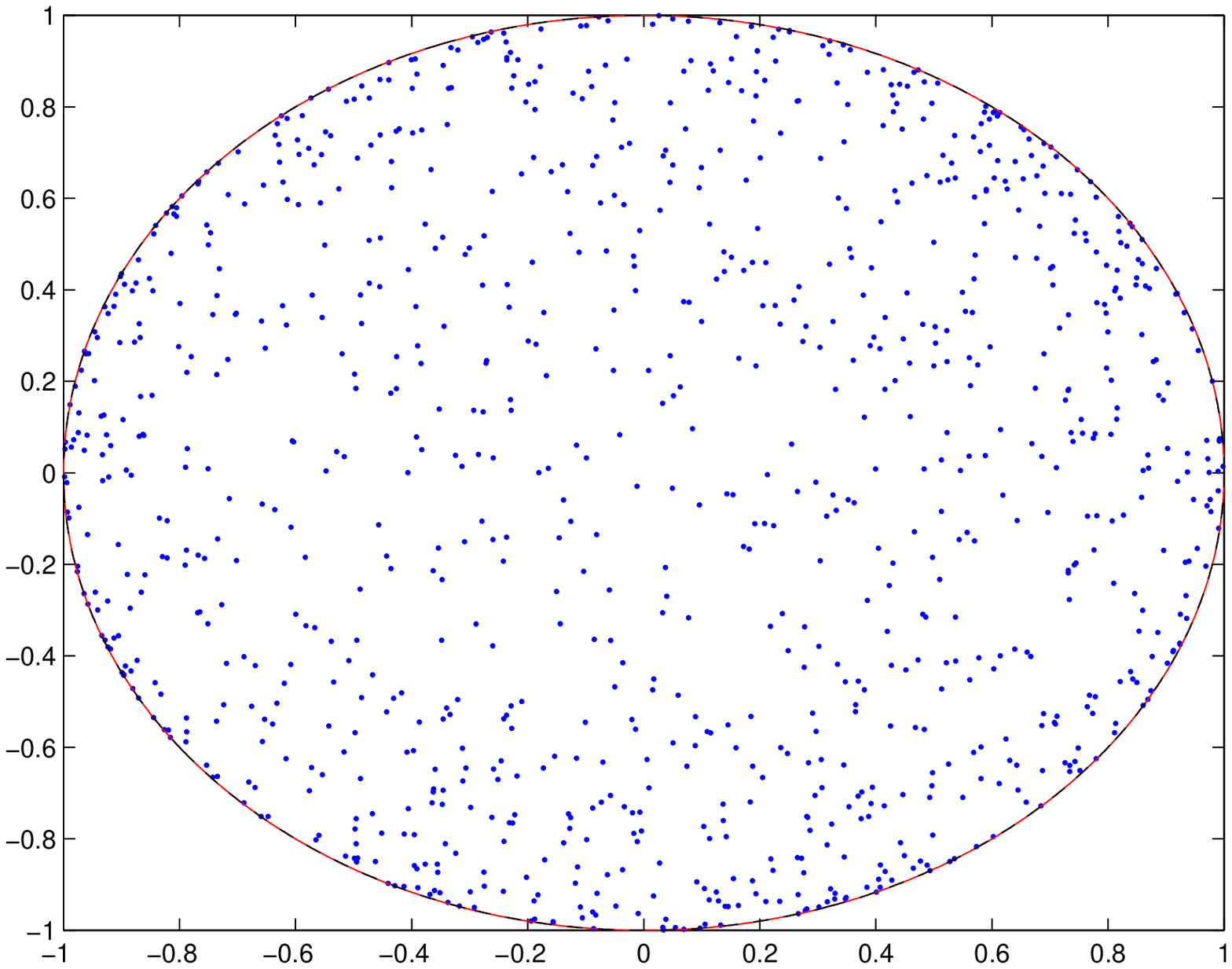}}
\hskip 1cm
\scalebox{0.22}[0.271]{\includegraphics{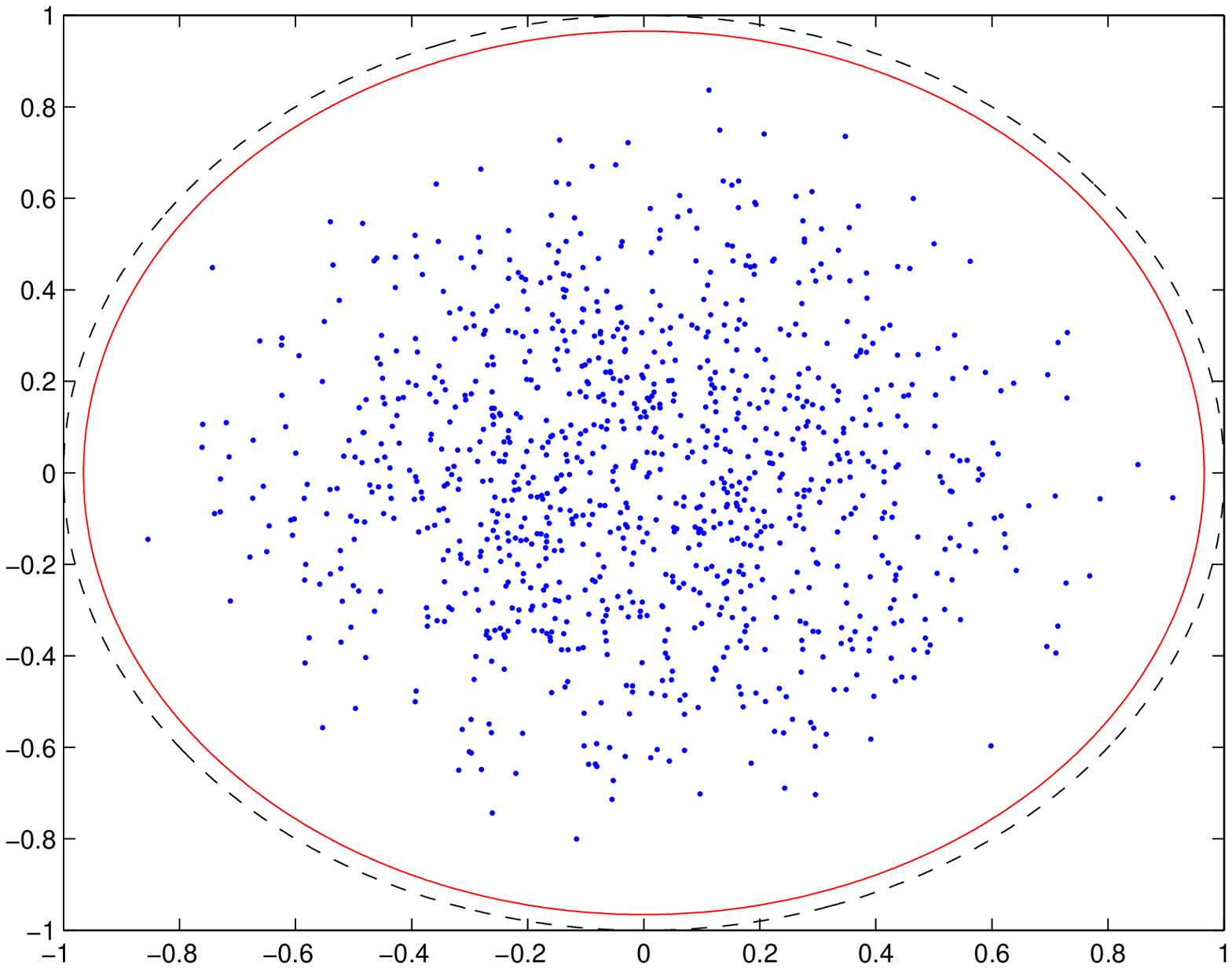}}}
\centerline{\scalebox{0.22}[0.271]{\includegraphics{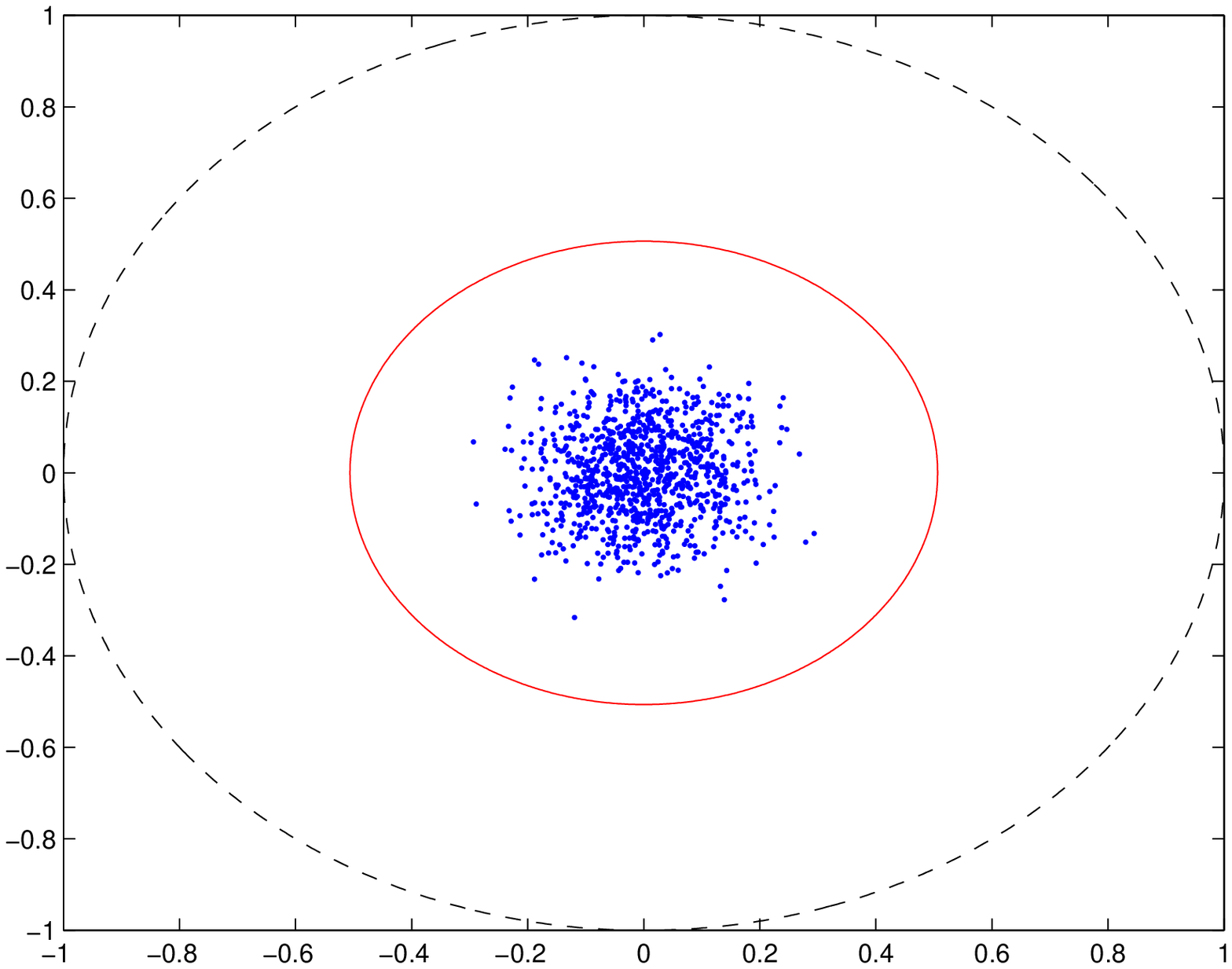}}
\hskip 1cm
\scalebox{0.22}[0.271]{\includegraphics{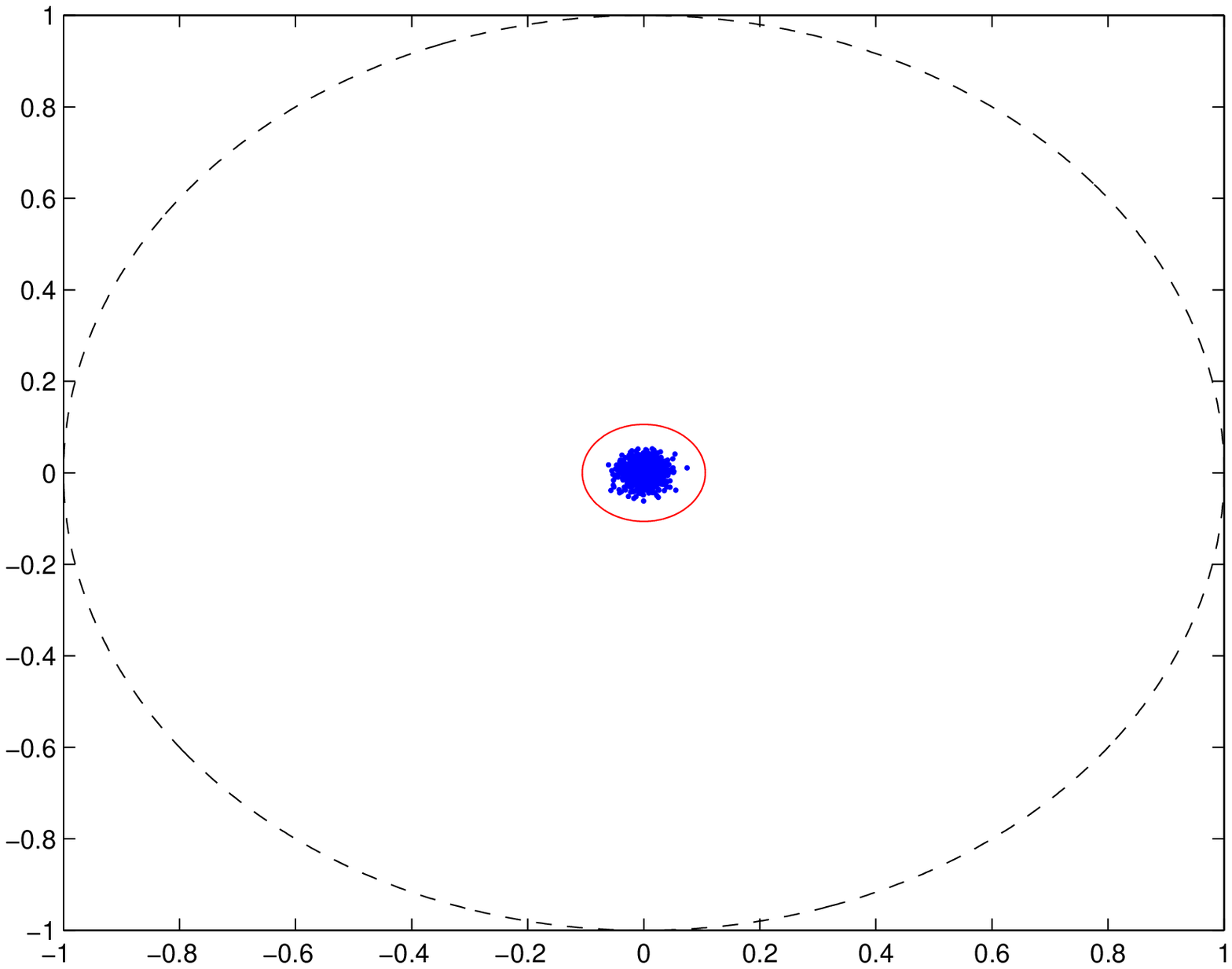}}}
\caption{Observable diameter of the sphere $\s^n$, $n=3,10,100,2500$.}
\label{fig:obs-diam}
\end{figure}

For the Euclidean spheres $\s^n$ of unit radius, equipped with the usual Euclidean distance and the (unique) rotation-invariant probability measure, one has, asymptotically as $n\to\infty$, ${\mathrm{CharSize}}(\s^n)\to \sqrt 2$, while ${\mathrm{ObsDiam}}(\s^n)=O(1/\sqrt n)$. 
Fig. \ref{fig:obs-diam} shows observable diameters (indicated by inner circles) corresponding to the threshold value $\kappa=10^{-10}$ of spheres $\s^n$ in dimensions $n=3,10,100,2500$, along with projections to the two-dimensional screen of randomly sampled 1000 points. 

Some other important examples of L\'evy families \cite{milman:86,L,gromov:99}
include: \\[1mm]
$\bullet$ Hamming cubes $\{0,1\}^n$ of two-bit $n$-strings equipped with the normalized Hamming distance $d(\sigma,\tau)=\frac 1n\sharp\{i\colon \sigma_i\neq\tau_i\}$ and the counting measure. The Law of Large Numbers is a particular consequence of
this fact, hence the name Geometric Law of Large Numbers sometimes used in
place of Concentration Phenomenon;\\[1mm]
$\bullet$ groups $SU(n)$ of special unitary $n\times n$ matrices, with the geodesic distance and Haar measure (unique invariant probability measure);\\[1mm]
$\bullet$  spaces $\R^d$ equipped with the Guassian measure with standard deviation $\sigma=1/\sqrt n$, \\[1mm]
$\bullet$ any family of expander graphs (\cite{gromov:99}, p. 197) with the normalized counting measure on the set of vertices and the path metric.

Any dataset whose observable diameter is small relative to the characteristic size will be suffering from dimensionality curse. For some recent work on this link in the context of data engineering, cf. \cite{francois} and references therein.

\subsection{Concentration function and separation distance}
One of many equivalent ways to reformulate the concentration phenomenon is this:
\begin{quote}
{\it for a typical ``high-dimensional'' structure $X$, if $A$ is a subset
containing at least half of all points, then the 
measure of the $\e$-neighbourhood
$A_\e$ of $A$ is overwhelmingly close to $1$ 
already for small values of $\e>0$. }
\end{quote}
More formally, one can prove that a family $(X_n,d_n,\mu_n)$ of $mm$-spaces is
a L\'evy family if and only if, whenever a Borel subset $A_n\subseteq X_n$ is picked up in every $X_n$ in such a way that $\mu_n(A_n)\geq 1/2$, 
one has $\mu_n((A_n)_\e)\to 1$ for every $\e>0$.
This reformulation allows to define
the most often used quantitative measure of concentration phenomenon,
the {\em concentration function}, $\alpha_X(\e)$, of an $mm$-space $(X,d,\mu)$, cf. \cite{milman:86,pestov:07}.
% we will not mention it here.
One sets $\alpha(0)=1/2$ and for all $\e>0$,
\[\alpha(\e)=1-\inf \left\{\mu(A_\e)\colon A\subseteq X,~~\mu(A)\geq \frac 12\right\},\]
where $A$ runs over Borel subsets of $X$. 
Clearly, a family of $mm$-spaces $(X_n)$ is L\'evy if and only if the concentration functions $\alpha_{X_n}(\e)$ converge to zero pointwise for all $\e>0$. 
 
Another such quantitative measure is
the {\em separation distance} \cite{gromov:99}. Let $\kappa>0$. The value ${\mathrm{sep}}_\kappa(X)$ of $\kappa$-separation distance of the $mm$-space $X$ is the supremum of all $\delta$ for which there are Borel sets $A,B\subseteq X$ at a distance $\geq\delta$ from each other which are both sufficiently large:
\[\mu(A)\geq\kappa,~~\mu(B)\geq\kappa.\]
\begin{figure}[ht]
\centering
\includegraphics[width=.5\textwidth]{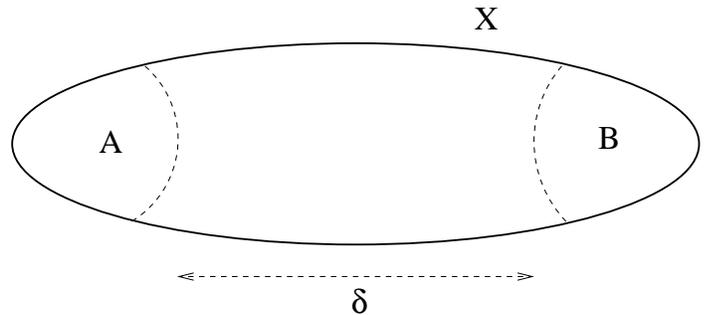}
\caption{To the notion of separation distance.}
\label{fig:sepdist}
\end{figure}
By setting in addition ${\mathrm{sep}}_0(X)={\mathrm{diam}}\,(X)$, one gets the {\em separation function} of $X$, ${\mathrm{sep}}\,(X)$, which is a non-increasing function from the interval $[0,1/2]$ to $\R$, vanishing at the right endpoint. 

\begin{figure}[ht]
\centering
\includegraphics[width=.5\textwidth]{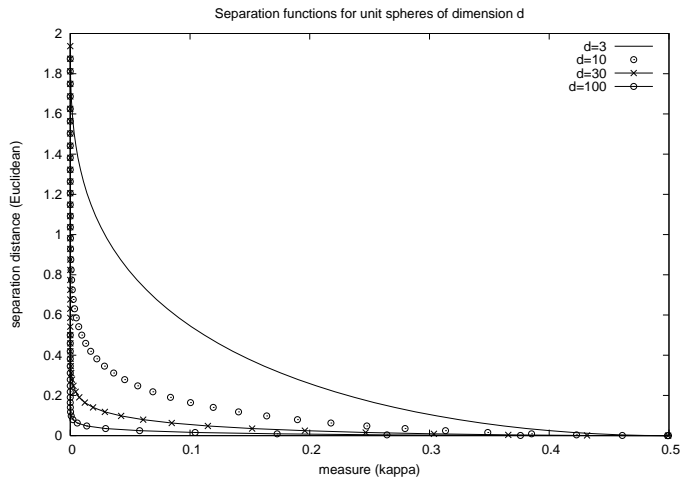}
\caption{Separation functions of the Euclidean spheres $\s^n$, $n=3,10,30,100$.}
\label{fig:sepfunctions}
\end{figure}

It is a simple exercise to verify that for all $\e,\kappa>0$
\[{\mathrm{sep}}_{\alpha(\e)}(X)\geq \e\mbox{ and }
\alpha_X({\mathrm{sep}_\kappa(X)}) \leq \kappa.\]
Thus, 
a family $(X_n)$ of $mm$-spaces is a L\'evy family if and only if ${\mathrm{sep}}_{\kappa}(X_n)$ converge to zero pointwise, cf. Fig. \ref{fig:sepfunctions}.

\subsection{Concentration and soft margin classification}

Here we will explain the concentration phenomenon in the language of soft margin classifiers. We will work in the setting of \cite{AB}, Subs. 9.2, 
assuming that the training dataset for a binary classification problem is modelled by a sequence of i.i.d. random variables distributed according to a probability measure $\nu$ on $Z=\Omega\times\{0,1\}$. Here $\Omega$ is the domain, in our case a metric space, and the classifying functions $f$ will be assumed $1$-Lipschitz. (For a detailed treatment of large margin classification problem for such functions, see \cite{vonlux:04}.)
The {\em margin} of a function $f\colon\Omega\to\{0,1\}$ on $(x,y)\in Z$ is defined as
\[{\mathrm{margin}}(f(x),y)=\left\vert f(x)-\frac 12\right\vert.\]
For a $\gamma\geq 0$ (margin parameter), define the {\em error} of $f$ with respect to $\nu$ and $\gamma$ as the probability
\[{\mathrm{er}}_\nu^\gamma(f)=\nu\left\{{\mathrm{margin}}(f(x),y)<\gamma\right\}.\]
The value $1-{\mathrm{er}}_\nu^\gamma(f)$ is a measure of how many datapoints admit a confident correct classification. 

\begin{theorem}
Let $(\Omega,d,\mu)$ be a metric space with measure, and let $\nu$ be a probability distribution on $Z=\Omega\times\{0,1\}$ with the marginals equal to $\mu$ on $\Omega$ and the Bernoulli distribution on $\{0,1\}$. Let $\gamma>0$. Then for every $1$-Lipschitz function $f$,
\[{\mathrm{er}}_\nu^\gamma(f)\geq 1 - 2\alpha(\gamma),\]
where $\alpha$ denotes the concentration function of the domain $(\Omega,d,\mu)$.
\label{th:gamma}
\end{theorem}

The result easily follows from the definition of the concentration function if one takes into account that the distribution $\nu$ induces a partition of $\Omega$ in two Borel subsets of measure $1/2$ each.
Conversely, one can bound the concentration function in terms of the uniform error: 
\[\alpha(\gamma)\leq 1-\sup_f{\mathrm{er}}_\nu^\gamma(f),\]
where the supremum is taken over all $1$-Lipschitz functions on $\Omega$. 

This formalizes the observation that in datasets suffering from dimensionality curse large margin classification with $1$-Lipschitz functions becomes impossible.

\section{\label{s:gromov}
Gromov's distance and concentration to a non-trivial space} 

\subsection{Definition}
Gromov's distance between two $mm$-spaces satisfies the usual axioms of a metric and is introduced in such a way that a family $(X_n)$ of $mm$-spaces forms a L\'evy family if it converges to a one-point space with regard to Gromov's distance. Thus, one can say that a dataset $X$ suffers from the curse of dimensionality if it is close to a one-point space in Gromov's distance. 
Intuitively, it means that the features of $X$ give away as little useful information about the intrinsic structure of $X$ as the features of the trivial one-point set, that is, not much more can be derived about $X$ from observations than about a one-point set. (For a formalization of this discussion in terms of Gromov's {\em observable diameter} of an $mm$-space, see \cite{pestov:07} and Gromov's original book \cite{gromov:99}.) 

Gromov's distance allows one to talk of {\em concentration to a non-trivial space}. 
In a sense, this is what happens in the context of principal manifold analysis, where one expects a dataset to concentrate to a low-dimensional manifold. 

Let $X=(X,d_X,\mu_X)$ and $Y=(Y,d_Y,\mu_Y)$ be two $mm$-spaces. The idea of Gromov's distance is that $X$ and $Y$ are close if every feature of $X$ can be matched against a similar feature of $Y$, and vice versa. For this purpose, one needs to represent all the features as functions on a common third space. This is achieved through a standard result in measure theory. Every $mm$-space $X$ can be {\em parametrized} by the unit interval: there is a measurable map $\phi\colon [0,1]\to X$ with the property that whenever $A\subseteq X$ is a Borel subset, one has
\[\mu_X(A)=\lambda(\phi^{-1}(A)),\]
where $\lambda$ is the Lebesgue measure on $[0,1]$ and $\phi^{-1}(A)=\{t\in [0,1]\colon \phi(t)\in A\}$ is the inverse image of $A$ under $\phi$. 

Introduce the distance ${\mathrm{me}}_1$ between measurable functions on $[0,1]$ as follows:
\[{\mathrm{me}}_1(f,g)=\inf\left\{\e>0\colon \lambda\{t\in [0,1]\colon \abs{f(t)-g(t)}> \e\}<\e\right\}.\]
This is indeed a metric, determining the well-known {\em convergence in measure}.

Now define the Gromov distance $d_{conc}(X,Y)$ between two $mm$-spaces $X$ and $Y$ as the infimum of all $\e>0$ for which there exist some suitable parametrizations $\phi_X$ and $\phi_Y$ of $X$ and of $Y$ respectively, with the following property. For every $f\in {\mathrm{Lip}}_1(X)$ there is a $g\in {\mathrm{Lip}}_1(Y)$ with
\begin{equation}
\label{eq:me1}
{\mathrm{me}}_1(f\circ \phi_X,g\circ \phi_Y)<\e,
\end{equation}
and vice versa: for every $g\in {\mathrm{Lip}}_1(Y)$ there is an $f\in {\mathrm{Lip}}_1(X)$ satisfying Eq. (\ref{eq:me1}).
\vskip .2cm

\begin{proposition}
Let $X$ be an $mm$-space. Then
\[\left(d_{conc}(X,\{\ast\})\leq\e/2\right)\Rightarrow \left(\alpha(\e)\leq\e/2\right)\Rightarrow \left(d_{conc}(X,\{\ast\})\leq \e\right).\]
\end{proposition}

\begin{pf} Suppose $d_{conc}(X,\{\ast\})\leq \e/2<1/2$ and let $A\subseteq X$, $\mu(A)\geq 1/2$. The distance function $d_A(x)=d(A,x)=\inf\{d(a,x)\colon a\in A\}$ is 1-Lipschitz and so differs from a suitable constant function $c$ by less than $\e/2$ on a set of measure $>1-\e/2$. Clearly, $c\leq\e/2$, and so $d_A$ can possibly take value $>\e$ on a set of measure $\leq\e/2$, meaning $\alpha(\e)\leq\e/2$. Conversely, if $d_{conc}(X,\{\ast\})\geq \e$, there exists a 1-Lipschitz function $f$ on $X$ which differs from its median value $M=M_f$ by at least $\e$ on a set of measure $\geq \e$. It means the existence of two sets, $A$ and $B$, such that $\mu(A)\geq 1/2$, $\mu(B)\geq \e/2$, and for all $a\in A$, $b\in B$ one has $\abs{f(a)-f(b)}\geq\e$, that is, $\alpha(\e)\geq \e/2$.
\end{pf}

\begin{figure}[ht]
\centering
\includegraphics[width=.5\textwidth]{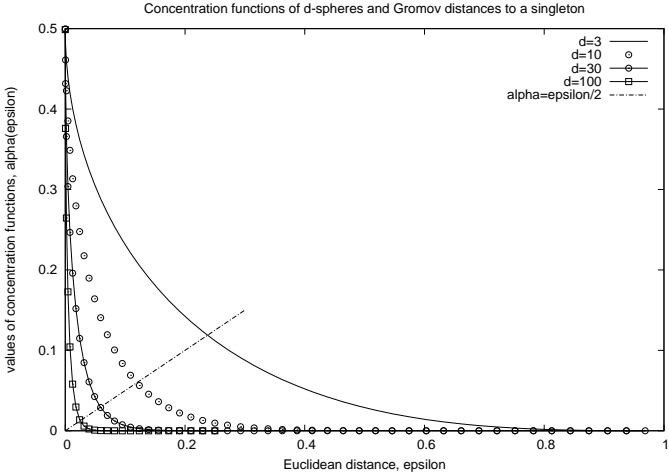}
\caption{Concentration functions of spheres $\s^n$, $n=3,10,30,100$, and the straight line $\alpha=\e/2$.}
\label{fig:distances}
\end{figure}

E.g. for the spheres the Gromov distance to a point is exactly the solution to the equation $\alpha_{\s^n}(\e)=\e/2$, Fig. \ref{fig:distances}.

\begin{corollary}
A family $(X_n)$ of $mm$-spaces is a L\'evy family if and only if it converges with regard to Gromov's distance to the trivial one-point space $\{\ast\}$.
\end{corollary}

\begin{remark}
\label{r:enough}
Notice that in the definition of Gromov's distance $d_{conc}(X,Y)$ one can replace throughout the sets ${\mathrm{Lip}}_1(X)$ of all $1$-Lipschitz functions with the sets of all 1-Lipschitz functions $f$ satisfying $\norm f_\infty\leq D$, where $D$ is an upper bound on the diameter of two metric spaces in question and $\norm{f}_\infty$ is the supremum norm of $f$.
\end{remark}

\subsection{Gromov and Monge-Kantorovich distances}
Let us compare the Gromov distance to the well-known {\em Monge-Kantorovich,} or {\em mass transportation,} distance (also known in computer science as the {\em Wasserstein}, or {\em Earth-mover's distance}), see \cite{rr:98}. Given two probability measures $\mu$ and $\nu$ on a metric space $(X,d)$, the mass transportation distance between them is
\[d_{mass}(\mu,\nu)=\inf_{\eta}\int_{X\times X} d(x,y)~d\eta,\]
where $\eta$ runs over all probability measures on $X\times X$ whose marginals are $\mu$ and $\nu$, respectively. Thinking of $\mu$ and $\nu$ as piles of sand of equal mass, $d_{mass}(\mu,\nu)$ is the smallest average distance that a grain of sand has to travel when the first pile is moved to take place of the second. 

\begin{proposition}
\label{p:mass}
$d_{conc}((X,d,\mu),(X,d,\nu))\leq \sqrt{d_{mass}(\mu,\nu)}$.
\end{proposition}

\begin{pf} Without loss in generality, one can assume both $\mu$ and $\nu$ to be non-atomic, and use the coordinate projections $\pi_i$, $i=1,2$, from the measure space $(X\times X,\eta)$ to parametrize the two $mm$-spaces in question. For every $f\in{\mathrm{Lip}}_1(X)$ the $L^1$-norm of the difference $\pi_i\circ f-\pi_2\circ f$ satisfies
\[\int_{X\times X}\abs{f(x)-f(y)}\,d\eta(x,y)\leq \int_{X\times X} d(x,y)~d\eta=d_{mass}(\mu,\nu),\]
whence the desired estimate follows easily. 
\end{pf}

No bound in the opposite direction is possible.
For instance, $d_{conc}(\s^n,\{0\})=o(1)$, while the mass transportation distance between the Haar measure on the sphere $\s^n$ and any Dirac point mass 
will be at least $1$. 

\subsection{Sampling}
If $(\Omega,d,\mu)$ is an $mm$-space and $X$ is a $\mu$-sample of $\Omega$, then $X$ becomes an $mm$-space on its own right if equipped with the restriction of the distance $d$ and the normalized counting measure. The following theorem states that random samples of an $mm$-space $\Omega$ will concentrate to it with  confidence approaching one as the sample size increases. 

Recall that a metric space $\Omega=(\Omega,d)$ is {\em totally bounded} if for every $u>0$ it can be covered with finitely many open balls of radius $u$, and the smallest such number, the {\em covering number}, is denoted $N(\Omega,u)$. For instance, every compact metric space is totally bounded.

\begin{theorem} 
\label{th:sampling}
Let $(\Omega,d,\mu)$ be a totally bounded metric space of diameter one equipped with a non-atomic Borel probability measure $\mu$. Let $\e,\delta>0$, and let $X$ be a random $\mu$-sample of $\Omega$ of size
\[n\geq \frac C{\e^4}\max\left\{\log\frac 2\delta,
\left(\int_{\e^2/8}^4\sqrt{N\left(\Omega,\frac u4\right)\log\left(\frac 4u+1 \right)} du \right)
\right\},\]
where $C>0$ is an absolute constant. \par
Then with confidence $>1-\delta$ one has:
\[d_{conc}(\Omega,X)<\e.\]
\end{theorem}

\begin{pf} The Rademacher averages of a class $\mathscr F$ of functions are capacity measures defined as follows: for every $n=1,2,3,\ldots$,
\begin{equation}
\label{eq:rademacher}
R_n({\mathscr F})={\mathbb E}_{\mu}{\mathbb E}_\e\left(\frac 1n\sup_{f\in{\mathscr F}}\left\vert \sum_{i=1}^n\e_i f(X_i) \right\vert \right),\end{equation}
where $X_i$, $i=1,2,\ldots,n$ are i.i.d. sample points according to the sample distribution $\mu$, and $\e_i$ are i.i.d. Rademacher random variables assuming equiprobable values $\pm 1$. 

Making use of Remark \ref{r:enough},
denote by ${\mathscr F}_{\Omega}$ the space of all $1$-Lipschitz functions on $\Omega$ with $\norm f_{\infty}\leq 1$, and similarly for ${\mathscr F}_X$. By Theorem 18 in \cite{vonlux:04}, for a suitable constant $C>0$,
\begin{equation}
\label{eq:bound}
R_n({\mathscr F}_{\Omega})\leq 2\e+ \frac C{\sqrt n} \int_{\e/4}^4
\sqrt{N\left(\Omega,\frac u4\right)\log\left(\frac 4u+1 \right)} du.
\end{equation}
Since the diameter and covering numbers of $X$ are majorized by those of $\Omega$, the inequality (\ref{eq:bound}) remains true if ${\mathscr F}_{\Omega}$ is replaced with ${\mathscr F}_{X}$. 

Because $(\Omega,\mu)$ is a standard Borel non-atomic probability space, it can be used instead of the unit interval to parametrize $X$ (with the normalized counting measure). Choose a Borel measurable parametrization $\phi\colon \Omega\to X$ with the property $\phi(x)=x$ for each $x\in X$. Let $\phi^\ast{\mathscr F}_X$ denote the set of all pull-back functions on $\Omega$ of the form $g=f\circ\phi$, $f\in{\mathscr F}_X$. Such functions are Borel measurable, though not necessarily Lipschitz. 
By the choice of $\phi$, the Rademachar averages of ${\mathscr F}_X$ and of $\phi^\ast{\mathscr F}_X$ coincide, and so Eq. (\ref{eq:bound}) continues to hold with $\phi^\ast{\mathscr F}_X$ in place of ${\mathscr F}_{\Omega}$.

Corollary 3 on p. 19 in \cite{mendelson:03}, applied to the function class ${\mathscr F}={\mathscr F}_{\Omega}$, together with the inequality (\ref{eq:bound}), implies that, under the condition 
\begin{equation}
\label{eq:mendelson}
n\geq \frac{C}{\e^2}\max\left\{\log\frac 1\delta, \int_{\e/4}^4
\sqrt{N\left(\Omega,\frac u4\right)\log\left(\frac 4u+1 \right)} du \right\},
\end{equation}
one has  
with confidence $>1-\delta$ that the empirical mean and the expected value of each $f\in{\mathscr F}={\mathscr F}_{\Omega}$ differ by less than $\e$:
\[\sup_{f\in{\mathscr F}}\left\vert \frac 1n\sum_{i=1}^n f(X_i) -{\mathbb E}_{\mu}f \right\vert <\e.\]
(Notice that in Eq. (\ref{eq:rademacher}) we use the normalization by $1/n$ as e.g. in \cite{vonlux:04}, while the normalization in \cite{mendelson:03} is by $1/\sqrt n$. Also, the constant $C$ in Eq. (\ref{eq:mendelson}) is different from that in Eq. (\ref{eq:bound}).) 

An analogous statement is true of the class ${\mathscr F}=\phi^\ast{\mathscr F}_X$. 
Consequently, under the assumption (\ref{eq:mendelson}), with confidence $>1-2\delta$, 
if $f\in {\mathscr F}_{\Omega}$ and $g\in \phi^\ast{\mathscr F}_X$ coincide on $X$, then $\norm{f-g}_1<2\e$, which, in its turn, easily implies ${\mathrm{me}}_1(f,g)< \sqrt{2\e}$. 

For every function $f\in{\mathscr F}_{\Omega}$ there is a function from the class $\phi^\ast{\mathscr F}_X$ taking the same values as $f$ at all points of $X$: this is the function $\left(f\vert_X\right)\circ\phi$. The converse is also true:
as is well known, every 1-Lipschitz function on a subspace of a metric space (e.g. $X$) admits an extension to a 1-Lipschitz function on the entire space (in our case, $\Omega$), cf. e.g. Lemma 7 in \cite{vonlux:04}.
if $g\in {\mathscr F}_X$, there exists a $\tilde g\in{\mathscr F}_{\Omega}$ extending $g$, and now $(g\circ\phi)\vert_X=\tilde g\vert_X$. We conclude: with confidence $>1-2\delta$, the Hausdorff distance between ${\mathscr F}_{\Omega}$ and the pull-back of ${\mathscr F}_X$ to $\Omega$ is bounded by $\sqrt{2\e}$. Therefore, 
\[d_{conc}(\Omega,X)<\sqrt{2\e}\]
with confidence $>1-2\delta$. Making a substitution $\e_{new} = \sqrt{2\e}$, $\delta_{new} = 2\delta$, we obtain the desired result.
\end{pf}

Since the above result is meant to be applied to low-dimensional manifolds, the values of the covering numbers are relatively low, and the theorem gives meaningful estimates for realistically sized sample sets.
For $\e=0.1$ and $\delta=10^{-6}$ they are on the order of thousands of points for $d=1$ (principal curves), tens of thousands for $d=2$ and millions for $d=3$. The estimates can be no doubt significantly improved. 

\section{\label{s:axiom}
An axiomatic approach to inner dimension}
\label{s:axiomatic}

\subsection{Axioms}

Let $\mathscr M$ denote some class of spaces with metric and measure (possibly including all of them), containing a family $(X_n)$ of spaces asymptotically approaching the $n$-dimensional unit Euclidean spheres $\s^n$ with their standard rotation-invariant probability measures:
\[d_{conc}(X_n,\s^n)\to 0.\]
These can be Euclidean cubes $[0,1]^n$ with the Lebesgue measure and the distance normalized by $1/\sqrt n$, the Hamming cubes $\{0,1\}^n$ with the normalized Hamming ($\ell^1$) distance and normalized counting measure, etc.

Let $\partial$ be a function defined for every member of $\mathscr M$ and assuming values in $[0,\infty)\cup\{\infty\}$. We call $\partial$ an {\em intrinsic dimension function} if it satisfies the following axioms:
\smallskip

\begin{enumerate}
\item ({\em \label{ax:conc}Axiom of concentration})
A family $(X_n)$ of members of $\mathscr M$ is a L\'evy family if and only if
$\partial(X_n)\uparrow\infty$.
\item ({\em Axiom of smooth dependence on datasets})  If $X_n,X\in{\mathscr M}$ and $d_{conc}(X_n,X)\to 0$, then $\partial(X_n)\to \partial(X)$.
\item ({\em Axiom of normalization})\footnote{Here recall that $f(n)=\Theta(g(n))$ if there exist constants $0<c<C$ and an $N$ with $c\abs{f(n)}\leq \abs{g(n)}\leq C\abs{f(n)}$ for all $n\geq N$. One says that the functions $f$ and $g$ asymptotically have the same order of magnitude.} For some (hence every) family $(X_n)\subseteq{\mathscr M}$ with the property $d_{conc}(X_n,\s^n)=o(1)$ one has $\partial(X_n)=\Theta(n)$.
\end{enumerate}
\smallskip

The first axiom formalizes a requirement that the intrinsic dimension is high if and only if a dataset suffers from the curse of dimensionality. The second axiom assures that a dataset $X$ well-approximated by a non-linear manifold $M$ has an intrinsic dimension close to that of $M$. The role of the third axiom is just to calibrate the values of the intrinsic dimension.

As explained in \cite{pestov:07}, the axioms lead to a paradoxical conclusion: every dimension function defined for all $mm$-spaces must assign to the trivial one-point space $\{\ast\}$ the value $+\infty$. This paradox is harmless and does not lead to any contradictions, furthermore one can avoid it is by restricting the class $\mathscr M$ to $mm$-spaces of a given characteristic size (i.e., the median value of distances between two points), which does not lead to any real loss in generality. 

\subsection{Dimension function based on separation distance} 

In \cite{pestov:07} we gave an example of a dimension function, the {\em concentration dimension} of $X$:
\begin{equation}\label{eq:concdim}
\dim_{\alpha}(X)=\frac{1}{\left[2\int_0^1 \alpha_X(\e)~d\e\right]^2}.\end{equation}
Here is another dimension function.

\begin{figure}[ht]
\centering
\includegraphics[width=.5\textwidth]{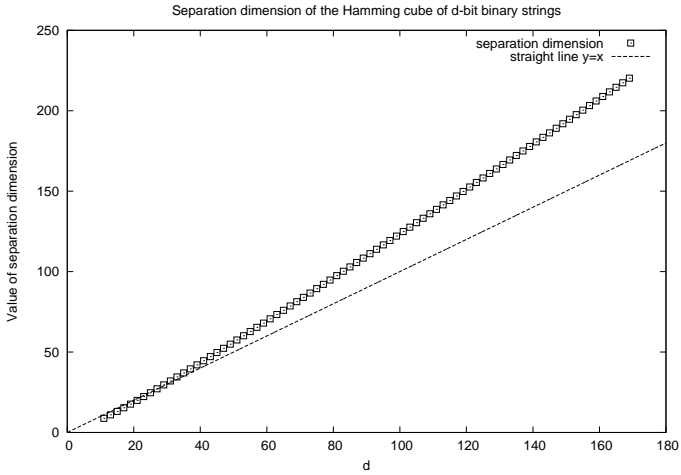}
\caption{Separation dimension $\partial_{sep}$ of the Hamming cube $\{0,1\}^d$, equipped with the normalized Hamming distance and normalized counting measure, $11\leq d\leq 169$, $d$ odd.}
\label{fig:sepDimHam}
\end{figure}

\begin{example} The quantity
\begin{equation}
\partial_{sep}(X)= \frac{d\kappa}
{\left[2\int_0^{\frac 12}{\mathrm{sep}}_{\kappa}(X)\,d\kappa\right]^2}
\end{equation}
defines an intrinsic dimension function on the class of all $mm$-spaces $X$ for which the above integral is proper (including, in particular, all spaces of bounded diameter). We call it the {\em separation dimension}. {\em (Cf. Fig. \ref{fig:sepDimHam}.)}
\end{example}

By judiciously choosing a normalizing constant, one can no doubt make the separation dimension of $\{0,1\}^n$ fit the values of $n$ much closer.

In fact, practically every concentration invariant from theory of $mm$-spaces leads to an example of an intrinsic dimension function, and the chapter 3$\frac 12$ of \cite{gromov:99} is a particularly rich source of such invariants.

\subsection{Dimension and sampling}

Most existing approaches to intrinsic dimension of a dataset have to confront the problem that, strictly speaking, the value of dimension of a finite dataset is zero, because it is a discrete object. 
On the contrary, as examplified by the Hamming cube $\{0,1\}^n$ (Fig. \ref{fig:sepDimHam}), our dimension functions make no difference between discrete and continuous $mm$-spaces. Moreover, the dimension of randomly sampled finite subsets approaches the dimension of the domain. The following is a consequence of Theorem \ref{th:sampling} and Axiom 2 of dimension function.

\begin{corollary} 
Let $\partial$ be a dimension function, and let $\Omega=(\Omega,d,\mu)$ be a non-atomic $mm$-space.
For every $\e>0$, $\delta>0$ there is a value $n_0=n_0(\partial,\Omega,\e,\delta)$ such that, whenever $X$ is a set of cardinality $\geq n_0$ randomly sampled from $\Omega$ with regard to the measure $\mu$, one has with confidence $>1-\delta$
\[\abs{\partial(\Omega)-\partial(X)}<\e.\]
\label{c:sampling}
\end{corollary}

Jointly with Theorems \ref{th:gamma} and \ref{th:sampling}, the above Corollary implies the following result which we state in a qualitative version.

\begin{corollary}
Let $\partial$ be a dimension function, and
let $(\Omega,d,\mu)$ be a non-atomic metric space with measure. Let $\nu$ be a probability distribution on $Z=\Omega\times\{0,1\}$ with the marginals equal to $\mu$ on $\Omega$ and the Bernoulli distribution on $\{0,1\}$. Then for every 
$\gamma,\delta,\e>0$ there are natural numbers $n_0$ and $d_0$ with the following property. Assume $\partial(\Omega,d,\mu)\geq d_0$. 
Let $n\geq n_0$ training datapoints be sampled from $Z$ according to the distribution $\nu$ an an i.i.d. fashion. Then with confidence $1-\delta$,
for every $1$-Lipschitz function $f$ the empirical error satisfies
\[{\mathrm{er}}_{\nu_n}^\gamma(f)\geq 1 - \e,\]
where $\nu_n$ is the empirical measure supported on the sample.
\end{corollary}

In other words, an intrinsically high-dimensional dataset does not admit large margin classifiers. 

\section{\label{s:complexity}
Complexity}

For the moment, we don't have any example of a dimension function that would be computationally feasible other than for well-understood geometrical objects (spheres, cubes...). 

\begin{theorem}
Fix a value $0<\kappa<1/2$. 
Determining the value ${\mathrm{sep}}_{\kappa}(X)$ of the separation function for finite metric spaces $X$ (with the normalized counting measure) is an $NP$-complete problem.
\end{theorem}

\begin{pf}
To a given finite metric space $X$ associate a graph with $X$ as the vertex set and two vertices $x,y$ being adjacent if and only if $d(x,y)\geq\kappa$. Now the problem of determining ${\mathrm{sep}}_{\kappa}(X)$ is equivalent to solving 
the largest balanced complete bipartite subgraph problem which is known to be $NP$-complete, cf. GT24 in \cite{gj:79}. 
\end{pf}

\section{\label{s:noise}
High-dimensional noise} 

Another deficiency of our approach in its present form is its sensitivity to noise.
We will consider an idealized situation where data is corrupted by high-dimensional Gaussian noise, as follows. Let $\mu$ be a probability measure on the Euclidean space $\R^d$. Assume that $\mu$ is supported on a compact submanifold $M$ of $\R^d$ of lower dimension $m\ll d$. If $\mu$ has density $p_\mu$ (that is, is absolutely continuous with regard to the Lebesgue measure), a dataset $X$ being sampled in the presence of gaussian noise means
\[X\sim p_\mu+ {\mathcal N}(0,\sigma^2),\]
where 
\[{\mathcal N}(0,\sigma^2) = \frac{1}{\sigma^n(2\pi)^{n/2}}e^{-\norm x^2/2\sigma^2}\]
is the density of the gaussian distribution $\gamma_n(0,\sigma^2)$. Equivalently, $X$ is sampled with regard to
the convolution of $\mu$ with the $d$-dimensional Gaussian measure:
\[Y\sim \mu\ast \gamma_n(0,\sigma^2),\]
in which form the assumption of absolute continuity of $\mu$ becomes superfluous. 
One can think of the $mm$-space $(\R^d,\mu\ast \gamma_n)$ with the Euclidean distance as a {\em corruption} of the original domain $(M,\mu)$. We will further assume that the amplitude of the corrupting noise is on the same order of magnitude as the size of $M$, that is, ${\mathbb{E}}_{\gamma_n}(\norm x)=O(1)$, or $\sigma=O(1/\sqrt d)$. 

Here is a result in the positive direction.

\begin{theorem} 
\label{th:corruption}
Let $M$ be a compact topological manifold supporting a probability measure $\mu$. Consider a family of embeddings of $M$ into the Euclidean space $\R^d$, $d\to\infty$ as a submanifold in such a way that the Euclidean covering numbers $N(M,\e)$, $\e>0$, grow as $o(d)$. Let $(M,\mu)$ be corrupted by the gaussian noise $\gamma_d(0,\sigma^2)$ of constant amplitude, that is, $\sigma=O(1/\sqrt d)$. Then the Gromov distance between the image of $(M,\mu)$ in $\R^d$ and its corruption by $\gamma_d(0,\sigma^2)$ tends to zero as $d\to\infty$. 
\end{theorem}

\begin{pf} 
For an $\e>0$, let $F$ be a finite $\e$-net for $M$. 
Denote by $\pi$ the orthogonal projection from $\R^d$ to the linear subspace $V$ spanned by $F$. Let $\pi_\ast\mu$ denote the push-forward of the measure $\mu$ to $V$, that is, for every Borel $A\subseteq V$ one has $(\pi_\ast\mu)(A)=\mu(\pi^{-1}(A))$. 

The mass transportation distance between $\mu$ and $\pi_\ast\mu$ is bounded by $\e$, and by Proposition \ref{p:mass} the Gromov distance between $M$ and $(V,\pi_\ast\mu)$ is bounded by $\e^{1/2}$. A similar argument gives the same upper bound for the Gromov distance between the gaussian corruption of $M$ and that of $(V,\pi_\ast\mu)$. 

The $mm$-space $(\R^d,(\pi_\ast\mu)\ast\gamma_n)$ can be parametrized by the identity mapping of itself (because the measure is non-atomic and has full support), while the projection $\pi$ parametrizes the space $(V,\pi_\ast\mu)$ by its very definition. If $f\in{\mathrm{Lip}}_1(V)$, then $\pi\circ f\in {\mathrm{Lip}}_1(\R^d)$. Conversely, let $f\in {\mathrm{Lip}}_1(\R^d)$. The fibers $\pi^{-1}(x)$, $x\in V$ are $(d-\abs F)$-dimensional affine subspaces, and the measure induced on each fiber by the measure $(\pi_\ast\mu)\ast\gamma_n$ approaches the gaussian measure $\gamma_{n-\abs F}(0,\sigma^2)$ with regard to the mass transportation distance as $d\to\infty$.
The function $\tilde f$ obtained from $f$ by integration over all fibers $\pi^{-1}(x)$, $x\in V$ belongs to ${\mathrm{Lip}}_1(V)$, and since $d-\abs F=O(d)$, the concentration of measure for gaussians (p. 140 in \cite{milman:86}) implies that for some absolute constant $C$, the functions $\pi\circ f$ and $f_\infty$ differ by less than $\e$ on a set of $(\pi_\ast\mu)\ast\gamma_d$-measure $>1-C\exp\left(-C\e^2/\sigma^2\right)$. In particular, if $d$ is large enough, the Gromov distance between $M$ and its gaussian corruption will not exceed $3\sqrt \e$, whence the result follows since $\e>0$ was arbitrary. 
\end{pf} 

\begin{corollary}
Under the assumptions of Theorem \ref{th:corruption}, the value of any dimension function $\partial$ for the corruption of $M$ converges to $\partial(M)$ as $d\to\infty$.
\end{corollary}

Unfortunately, this result does not extend to finite samples, because the required size of a random sample of $M$ in the presence of noise is unrealistically high: the covering numbers of $(\R^d,\mu\ast\gamma_d)$ go to infinity exponentially fast (in $d$), and Theorem \ref{th:sampling} becomes useless.

As an illustration, consider the simplest case possible.

\begin{proposition}
Let $M=\{0\}$ be a singular one-point manifold, and let $X$ be sampled from $M$ in the presence of gaussian random noise of moderate amplitude, that is,
\[X\sim {\mathcal N}(0,\sigma^2),\]
where $\sigma^2=1/d$. Assume the cardinality of the sample $X$ to be constant, $\abs X=O(1)$. Then the Gromov distance between $M=\{0\}$ and $X$ tends to a positive constant ($\sqrt 2/3\approx 0.47$) as $d\to\infty$. 
\end{proposition}

\begin{pf}
It is a well-known manifestation of the curse of dimensionality that, as $d\to\infty$, the distances between pairs of points of $X$ strongly concentrate near the median value, which in this case will tend to $\sqrt 2$.
Thus, a typical random sample $X$ will form, for all practical purposes, a discrete metric space of diameter $\sim \sqrt 2$.
In particular, ${\mathrm{Lip}}_1(X)$ will contain numerous $1$-Lipschitz functions that are highly non-constant, and the Gromov distance from $X$ to the one-point space $M$ is seen to tend to the value $\sqrt 2/3$.
\end{pf}

For manageable sample sizes (up to millions of points) the above will already happen in moderate to high dimensions. 

\begin{example} 
For $d=50$, a random sample $X$ as above of $s= 10^6$ points will contain, with confidence $> 0.99$, a $1$-separated subset $S$ containing $\geq 95$ \% of all points (that is, every two points of $S$ are at a distance $\geq 1$ from each other). Consequently, ${\mathrm{sep}}_{0.475}(X)\geq 1$, and
the separation dimension $\partial_{sep}(X)$ will not exceed $1.125$. (At the same time, $\partial_{sep}(\{0\})=+\infty$.)
\end{example}

We conclude: the proposed intrinsic dimension of discrete datasets of realistic size is unstable under random high-dimensional noise of moderate amplitude. 

\section{\label{s:comparison}
Comparison to other approaches}

\subsection{The intrinsic dimensionality of Ch\'avez et al.}

The following interesting version of intrinsic dimension was proposed by Ch\'avez {\em et al.} \cite{chavez:01} who called it simply {\em intrinsic dimensionality}. 
Let $(X,d,\mu)$ be a space with metric and measure.
Denote by $m(d)$ the mean of the distance function $d\colon X\times X\to\R$ on the space $X\times X$ with the product measure. Assume $m(d)<\infty$. 
Let $\sigma(d)$ be the standard deviation of the same function. The intrinsic dimensionality of $X$ is defined as
\begin{equation}
\dim_{dist}(X)=\frac{m^2(d)}{2\sigma^2(d)}.
\label{eq:int}
\end{equation}

\begin{theorem}
The intrinsic dimensionality satisfies:
\begin{itemize}
\item a weaker version of Axiom 1: if $(X_n,d_n,\mu_n)$ is a L\'evy family of spaces 
with bounded metrics, then $\dim_{dist}(X_n,\{\ast\})\to \infty$,
\item a weaker version of Axiom 2: if\hfill $d_{conc}(X_n,X)\to 0$ and $m(d_n)\to m(d)$, then $\dim_{dist}(X_n)\to \dim_{dist}(X)$,
\item Axiom 3.
\end{itemize}
\label{th:chavez}
\end{theorem}

For a proof, as well as a more detailed discussion, see \cite{pestov:07}, where in particular it is shown on a number of examples that the dimension Ch\'avez {\em et al.} and our dimension can behave in quite different ways between themselves (and of course from the topological dimension).

\subsection{Some other approaches}
The approaches to intrinsic dimension listed below are all quite different both from our approach and from that of Ch\'avez {\em et al.}, in that they are set to emulate various versions of {\em topological} (i.e. essentially local) dimension. In particular,
all of them fail both our Axioms 1 and 2.

$\bullet$ {\em Correlation dimension,} which is a computationally efficient version of the box-counting dimension, see \cite{cv,tmgm}.

$\bullet$ {\em Packing dimension}, or rather its computable version as proposed and explored in \cite{Kegl}.

$\bullet$ {\em Distance exponent} \cite{ttf}, which is a version of the well-known Minkowski dimension.

$\bullet$ An algorithm for estimating the intrinsic dimension based on the Takens theorem from differential geometry \cite{pa}.

$\bullet$
A non-local approach to intrinsic dimension estimation based on entropy-theoretic results is proposed in \cite{ch}, however in case of  manifolds the algorithm will still return the topological dimension, so the same conclusions apply.

\section{Discussion}
\label{s:discussion}

We have proposed a new concept of the intrinsic dimension of a dataset or, more generally, of a metric space equipped with a probability measure. Dimension functions of the new type behave in a very different way from the more traditional approaches, and are closer in spirit to, though still different from, the notion put forward in \cite{chavez:01} (cf. a comparative discussion in \cite{pestov:07}). In particular, high intrinsic dimension indicates the presence of the curse of dimensionality, while lower dimension expresses the existence of a small set of well-dissipating features and a possibility of dimension reduction of $X$ to a low-dimensional feature space. The intrinsic dimension of a random sample of a manifold is close to that of the manifold itself, and for standard geometric objects such as spheres or cubes the values returned by our dimension are ``correct''.

Two main problems pinpointed in this article are prohibitively high computational complexity of the new concepts, as well as their
instability under random high-dimensional noise.

The root cause of both problems is essentially the same: the class of all $1$-Lipschitz functions is just too broad to serve as the set of admissible features. The richness of the spaces ${\mathrm{Lip}}_1(X)$ explains why computing concentration invariants of an $mm$-space is hard: roughly speaking, there are just too many feature functions on the space that are to be examined one by one. The abundance of Lipschitz functions on a discrete metric space $X$ is exactly what makes the Gromov distance from a random gaussian sample to a manifold large.  

At the same time, there is clearly no point in taking into account, as a potential feature, say, a typical polynomial function of degree $10$ on the ambient space $\R^{100}$, because such a function may contain up to ${100\choose 10}\sim 1.7\times 10^{13}$ monomials. Since we cannot store, let even compute, such a function, why should we care of it at all? 

A way out, as we see it, consists in refining the approach and modelling a dataset as a pair $(X,{\mathscr F})$, consisting of an $mm$-space $X$ {\em together with a class of admissible features,} $\mathscr F\subseteq {\mathrm{Lip}}_1(X)$, whose statistical learning capacity measures (VC-dimension, covering numbers, Rademacher averages, etc.) are limited. This will accurately reflect the fact that in practice one only uses features that are computationally cheap, and will allow a systematic use of Vapnik-Chervonenkis theory.

All the main concepts of asymptotic geometric analysis will have to be rewritten in the new framework, and this seems to be a potentially rewarding subject for further investigation. A theoretical challenge would to be obtain noise stability results under the general statistical assumptions of \cite{blanchard:06}.

Finally, the Gromov distance between two $mm$-spaces, $X$ and $Y$, is determined on the basis of comparing the features of $X$ and $Y$ rather than the spaces themselves, which opens a possibility to try and construct an approximating principal manifold to $X$ by methods of unsupervised machine learning by optimizing over suitable sets of Lipschitz functions, as in \cite{vonlux:04}.

The concept of dimension in mathematics admits a very rich spectrum of interpretations. We feel that the topological versions of dimension have been dominating applications in computing to the detriment of other approaches. We feel that the concept of dimension based on the viewpoint of asymptotic geometric analysis could be highly relevant to analysis of large sets of data, and we consider this article as a small step in the direction of developing this approach. 

\section*{Acknowledgements}
The author is grateful to three anonymous referees of this paper for a number of suggested improvements. Research was supported by NSERC discovery grant and University of Ottawa internal grants.

\bibliographystyle{alpha}

\begin{thebibliography}{}

\bibitem{AB} Anthony, M. \& Bartlett, P. (1999). Neural network learning: theoretical foundations. Cambridge:Cambridge University Press. 

\bibitem{BGRS}
Beyer, K., Goldstein, J., Ramakrishnan, R., \& Shaft, U. (1999).
When is ``nearest neighbor'' meaningful? In Proc. 7-th Intern.
Conf. on Database Theory
(ICDT-99), Jerusalem (217--235).

\bibitem
{blanchard:06}
Blanchard, G., Kawanabe, M., Sugiyama, M., Spokoiny, V., \& M\"uller, K.-R. (2006).
\newblock
In search of non-Gaussian component of a high-dimensional distribution.
\newblock
Journal of Machine Learning Research, 7, 247--282.

\bibitem{cv}
Camastra, F. \& Vinciarelli, A. (2002). Estimating the intrinsic dimension of data 
with a fractal-based method. IEEE Transactions on Pattern Analysis and 
Machine Intelligence, 24 (10), 1404-1407.

\bibitem 
{chavez:01}
Ch\'avez, E., Navarro, G. Baeza-Yates, R. \& Marroqu\'\i n, J. L. (2001).
\newblock
Searching in metric spaces.
\newblock ACM Computing Surveys, 33, 273--321.

\bibitem{ch} Costa, J.  and Hero, A. O. (2004). Geodesic entropic graphs for dimension and entropy estimation in manifold learning. IEEE Trans. on Signal Process., 52(8), 2210-2221.

\bibitem{francois}
Fran\c cois, D., Wertz, V., \& Verleysen, M. (2007).
The concentration of fractional distances.
IEEE Trans. on Knowledge and Data Engineering 19(7), 873--886.

\bibitem 
{gj:79}
Garey, M.R. \& Johnson, D.S. (1979).
\newblock Computers and Intractability, A Guide to the Theory of NP-completeness. 
\newblock San Francisco:Freeman.

\bibitem 
{gromov:99} 
Gromov, M. (1999). \newblock
Metric Structures for Riemannian and
Non-Riemannian Spaces.  
\newblock Progress in Mathematics \textbf{152}. Birkhauser
Verlag.

\bibitem 
{hein:05}
Hein, M., \& Audibert, J.-Y. (2005).
\newblock
Intrinsic dimensionality estimation of submanifolds in $\R^d$.
\newblock
In: L. de Raedt and S. Wrobel (eds.), Proc. 22nd Intern. Conf. on Machine Learning (ICML) (pp. 289--296), AMC Press. 

\bibitem 
{hein:07}
Hein, M. \& Maier, M. (2007).
\newblock
Manifold denoising as preprocessing for finding natural representations of data.
\newblock
In Proc. Twenty-Second AAAI Conference on Artificial Intelligence
(Vancouver, B.C.),  pp. 1646--1649.

\bibitem{Kegl}
K\'egl, B. (2003).  Intrinsic dimension estimation using packing numbers. In Advances in Neural Information Processing Systems [NIPS 2002, Vancouver, B.C., Canada], vol~15 (pp.~681--688), The MIT Press.

\bibitem{L}
Ledoux, M. (2001). The concentration of measure phenomenon. Math. Surveys and 
Monographs {\bf 89}, Providence: Amer. Math. Soc.

\bibitem 
{mendelson:03}
Mendelson, S. (2003). 
\newblock
A few notes on statistical learning theory.
\newblock
In S. Mendelson, A.J. Smola (Eds), Advanced Lectures in Machine Learning, Lect. Notes in Computer Sci. 2600 (pp. 1-40), Springer.

\bibitem 
{milman:00} Milman, V. (2000). \newblock
Topics in asymptotic geometric analysis. In
\newblock  
Geometric and Functional Analysis, special volume GAFA2000 (pp. 792--815).

\bibitem 
{milman:86}                                                   
Milman, V.D. and Schechtman, G. (1986). \newblock
Asymptotic Theory of Finite-Dimensional 
Normed Spaces (with an Appendix by M. Gromov).                           
Lecture Notes in Math. \textbf{1200}, Springer.

\bibitem 
{pestov:00}
Pestov, V. (2000). 
\newblock On the geometry of similarity search: dimensionality curse and concentration of measure.
\newblock {\em Inform. Process. Lett.} 73, 47--51.

\bibitem 
{pestov:07} 
Pestov, V. (2007).
\newblock Intrinsic dimension of a dataset: what properties does one expect?
\newblock In: Proc. of the 22-nd Int. Joint Conf. on Neural Networks (IJCNN'07), Orlando, FL (pp. 1775--1780). 

\bibitem{pa} Potapov, A. \& Ali, M.K. (2002).
Neural networks for estimating intrinsic dimension.
    Phys. Rev. E  65 (2a) (4), 046212.1-046212.7.
    
\bibitem 
{rr:98} 
Rachev, S.T. \& R\"uschendorf, L. (1998).
\newblock
Mass Transportation Problems. Volume I: Theory. Volume II:
Applications. 
\newblock NY--Berlin--Heidelberg:Springer.

\bibitem{tmgm} Tatti, N., Mielikainen, T., Gionis, A. \& Mannila, H. (2006). What is the dimension of 
    your binary data? In 6th International Conference on Data Mining (ICDM), Hong Kong (pp. 603-612).
    
\bibitem{ttf} 
Traina, C., Jr., Traina, A.J.M. \& Faloutsos, C. (1999).
Distance exponent: A new concept for selectivity estimation in metric
trees. Technical Report CMU-CS-99-110, Computer Science Department, 
Carnegie Mellon University.
    
\bibitem 
{vonlux:04}
von Luxburg, U. and Bousquet, O. (2004).
\newblock Distance-based classification with Lipschitz functions.
\newblock Journal of Machine Learning Research 5, 669--695.

\end{thebibliography}

\end{document}